\documentclass[pra,showpacs,twocolumn]{revtex4-1}

\usepackage[utf8]{inputenc}
\usepackage[english]{babel}
\usepackage[T1]{fontenc}

\usepackage{graphicx}
\usepackage{amssymb,amsfonts,amsmath}
\RequirePackage[colorlinks=true,linkcolor=blue]{hyperref}

\newcommand{\Tr}[0]{\operatorname{Tr}}  
\newcommand{\diag}[0]{\mathrm{diag}}  
\newcommand{\ie}[0]{\textit{i.e.}}  
\newcommand{\eg}[0]{\textit{e.g.}}  
\renewcommand{\P}[0]{\mathcal{P}}  
\newcommand{\D}[0]{\mathcal{D}}  
\newcommand{\U}[0]{\mathcal{U}}  
\newcommand{\SSKF}[0]{\mathrm{SSKF}} 
\newcommand{\vN}[0]{\mathrm{vN}}  
\newcommand{\HS}[0]{\mathrm{HS}}  
\newcommand{\EDPW}[0]{\mathrm{EDPW}}  
\newcommand{\lin}[0]{\mathrm{lin}}  

\begin{document}

\title{Polarization monotones of two-dimensional and three-dimensional random electromagnetic fields}

\author{G.M. Bosyk}
\affiliation{%
 Instituto de Física La Plata, UNLP, CONICET, Facultad de Ciencias Exactas, 1900 La Plata, Argentina}
\author{G. Bellomo}%
 \email{Corresponding author: gbellomo@fisica.unlp.edu.ar}
\affiliation{%
CONICET-Universidad de Buenos Aires, Instituto de Investigación en Ciencias de la Computación (ICC), Buenos Aires, Argentina}
\author{A. Luis}
\affiliation{%
 Departamento de Óptica, Facultad de Ciencias Físicas,\\ Universidad Complutense, 28040 Madrid, Spain}




\begin{abstract}
We propose a formal resource-theoretic approach to quantify the degree of polarization of two and three-dimensional random electromagnetic fields. This endows the space of spectral polarization matrices with the orders induced by majorization or convex mixing that naturally recover the best-known polarization measures.
\end{abstract}


\date{\today}

\maketitle

\section{Introduction} \label{sec:introduction}
The problem of how to extend the notion of degree of polarization from two to three-dimensional random electromagnetic (EM) fields is recurrent in the literature of classical statistical optics, mainly because there is no agreement within the optics community on the definition of the notion of unpolarized states for the latter case (see, \eg,~\cite{Setala2002a,Luis2005a,Ellis2005a,Ellis2005b,Ellis2005c}). As a consequence, several measures of the degree of polarization have been introduced, many of them leading to contradictory assertions. Here, we approach this controversy by appealing to the resource-theoretic formalism originally introduced for entanglement and quantum coherence (see, \eg,~\cite{Vidal2000,Baumgratz2014}). To show its adequateness, we first analyze the two-dimensional (2D) case. Then, we test the formalism constructing the corresponding resource theories that arise when following the two different claims most commonly encountered in the literature about what an unpolarized state is in the three-dimensional (3D) case. Our presentation is based on the idea that any \textit{bonafide} degree of polarization must either decrease or stay constant under the nonpolarizing operations to be defined by the corresponding resource theory, thus providing a firm theoretical basis to support physical intuition.

The outline of the work is as follows. In Sec.~\ref{sec:polarization}, we introduce the set of polarization states for which we develop the resource theories of polarization. Our proposal and main results are given in Sec.~\ref{sec:resource}. In Sec.~\ref{subsec:formalism}, we provide the general framework for any resource theory of polarization. In Sec.~\ref{subsec:2D}, we present the  resource theory of polarization for 2D EM fields. In Sec.~\ref{subsec:3D}, we develop two resource theories for 3D EM fields: one based on a majorization partial order (Sec.~\ref{subsub:majorization}), the other one based on a convex mixing preorder (Sec.~\ref{subsub:convex}); and we classify the most well-known measures of degree of polarization present in the literature, in their corresponding theory. In addition, we compare both resource theories (Sec.~\ref{subsub:comparison}). Finally some conclusions are drawn in Sec.~\ref{sec:conclusions}.

\section{Polarization density matrix and polarization space}
\label{sec:polarization}
When dealing with statistically stationary random $d$-dimensional electromagnetic fields, the polarization properties at point $\vec{r}$ and frequency $\omega$ can be described by the $d \times d$ spectral polarization matrix, $\Phi(\vec{r},\omega)$, with entries
\begin{equation}\label{eq:polmatrixetries}
  \Phi_{i,j}(\vec{r},\omega) = \langle E_i(\vec{r},\omega) E^*_j(\vec{r},\omega) \rangle,
\end{equation}
where $i,j = 1, \ldots, d$ (see, \eg,~\cite{MandelBook}). Here, $E_i(\vec{r},\omega)$ denotes the $i$ component of a single realization of the electric field, whereas the angle brackets and asterisk denote ensemble averaging and complex conjugation, respectively. Hereafter, we omit the explicit dependence on $\vec{r}$ and $\omega$ of the quantities derived from the polarization matrix. Let us introduce the normalized version of $\Phi$, that is,
\begin{equation}\label{eq:densitymatrix}
  \rho = \frac{\Phi}{\Tr \Phi},
\end{equation}
where $\Tr$ denotes the trace operation ($\Tr\Phi$ accounts for the total intensity of the electric field~\cite{Gil2014}). This normalized version of $\Phi$ can be seen as a quantum density operator: a trace-one and semidefinite positive operator. By analogy to the quantum case, we call $\rho$ the polarization density matrix or state of polarization. Also, let us define the set of density matrices, $\D = \{\rho \in \mathbb{C}^{d \times d}: \rho \geq 0 \ \text{and} \ \Tr\rho =1\}$. In its diagonal basis, $\rho$ takes the form $\rho = \sum_{i=1}^{d}\rho_ i \rho^{i\mathrm{P}}$, where $\{\rho_ i\}$ are the eigenvalues and $\rho^{i\mathrm{P}}$ is the null matrix with the $(i,i)$ component equal to $1$. This decomposition can be interpreted as an incoherent mixture of orthogonal uncorrelated maximally polarized states.

As it has been remarked in the literature, the degree of polarization is a basis-independent property for any dimension (see, \eg,~\cite{Gamel2014}). As a consequence, we are interested in measures that depend only on the eigenvalues of $\rho$ and we can restrict the domain of interest introducing the equivalence relation $\sigma \sim \rho$ if and only if $\sigma = U \rho U^\dag$, where $U$ is an arbitrary unitary matrix. In this way, the equivalence class is $[\rho] = \{\sigma \in \D :  \rho \sim \sigma \}$, and the quotient set of polarization, $\D/\sim$, is given by the set of all equivalence classes. Without loss of generality, we can sort the eigenvalues of $\rho$ in a nonincreasing order and define the polarization space by the convex set $\P = \{\rho \in  \D/\sim : \rho = \diag(\rho_1,\ldots, \rho_d)$, with $\sum_{i=1}^{d} \rho_i = 1$ and $\rho_i \geq \rho_{i+1} \geq 0\}$. This set has already been introduced in different geometric approaches to polarization (see, \eg,~\cite{Saastamoinen2004,Sheppard2012,Gamel2014}).

\begin{figure}[htbp]
 \centering
 \includegraphics[width=\linewidth]{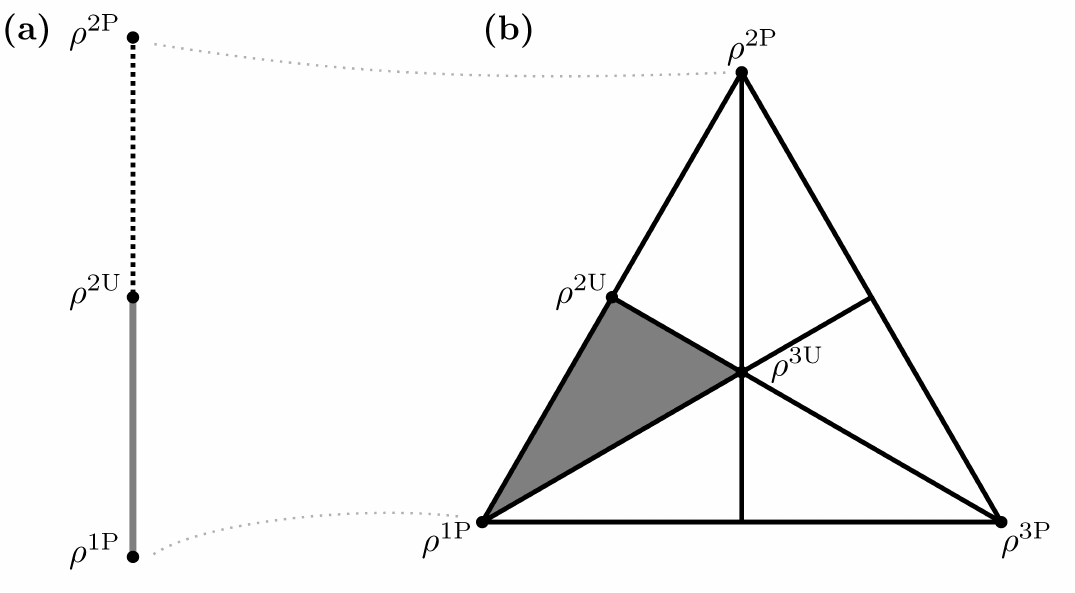}
 \caption{Polarization space representation. (a) For $d=2$ (solid line), $\P$ is embedded in the 1-simplex defined by $\rho^{\mathrm{1P}}$ and $\rho^{\mathrm{2P}}$. (b) For $d=3$ (gray triangle), $\P$ is embedded in the 2-simplex formed by the convex hull of $\rho^{\mathrm{1P}}$, $\rho^{\mathrm{2P}}$, and $\rho^{\mathrm{3P}}$.
 \label{fig:polspace}}
\end{figure}

In what follows, we restrict the analysis to the two- and three-dimensional cases. For $d=2$, $\P$ can be geometrically represented by a segment whose vertices are given by $\rho^{\mathrm{1P}} = \diag(1,0)$ and $\rho^{\mathrm{2U}} = \frac{1}{2}\diag(1,1)$, which is embedded in the 1-simplex [see Fig.\ref{fig:polspace}(a)]. Accordingly, any $2 \times 2$ polarization density matrix can be written as a convex combination of these extreme points, that is,
\begin{equation}
    \rho = (\rho_1 - \rho_2) \, \rho^{\mathrm{1P}} +  2 \, \rho_2 \, \rho^{\mathrm{2U}}.
    \label{eq:dp2D}
\end{equation}
For $d=3$, $\P$ is represented by a triangle with vertices $\rho^{\mathrm{1P}} = \diag(1,0,0)$, $\rho^{\mathrm{2U}} = \frac{1}{2}\diag(1,1,0)$, and $\rho^{\mathrm{3U}} = \frac{1}{3}\diag(1,1,1)$, which is embedded in the 2-simplex [see Fig.\ref{fig:polspace}(b)]. As a consequence, an arbitrary $3 \times 3$ polarization density matrix can be written as a convex combination of these extreme points, that is,
\begin{equation}
    \rho = (\rho_1 - \rho_2) \, \rho^{\mathrm{1P}} + 2\,(\rho_2 - \rho_3) \, \rho^{\mathrm{2U}} + 3\, \rho_3 \, \rho^{\mathrm{3U}}.
    \label{eq:dp3D}
\end{equation}
This decomposition has been also considered in~\cite{Gil2007,Gil2016}.

\section{Resource-theoretic approach}
\label{sec:resource}

\subsection{Formalism for a resource theory of polarization}
\label{subsec:formalism}
A formal resource theory for polarization has to be built from the following basic components: (i) the unpolarized states, (ii) a set of nonpolarizing operations, and (iii) the polarized states. These three concepts are not independent of each other. Indeed, the nonpolarizing operations must not generate polarized states from unpolarized ones. Therefore, any assumption made about one of these ingredients has an effect on the others. The idea is to first define the notion of being unpolarized. Then, the nonpolarizing operations are introduced as those that leave invariant the set of unpolarized states. More precisely, a resource theory for polarization must be such that
\begin{itemize}
  \item[(i)] there exists a set $\U$ of unpolarized states;
  \item[(ii)] there exists a class of nonpolarizing operations $\Lambda$ that preserves the set $\U$, \ie, $\Lambda(\rho) \in \U$ for all $\rho \in \U$.
\end{itemize}
So far, the only distinction among the states is to be unpolarized or not: given a state $\rho$, then either $\rho \in \U$ or $\rho \not\in \U$. In order to get some hierarchy among the polarized states, we must determine the nonpolarizing operations. Let us note that while the nonpolarizing operations can, in principle, convert a polarized state into another one, our intuition says that these operations can not convert one polarized state into another with greater degree of polarization. Such ``intuition'' has a status of definition in the theory: we postulate that any well defined measure of the degree of polarization must satisfy a monotonic nonincreasing behavior under the action of the nonpolarizing operations. More precisely, a \textit{bona fide} measure of the degree of polarization of $\rho$, $P(\rho): \P \rightarrow \mathbb{R}$, must be such that
\begin{equation}\label{eq:montonia}
  P(\Lambda(\rho)) \leq P(\rho), \quad \forall \, \rho \in \P.
\end{equation}
Thus, the intuition that the $\Lambda$ operations do not increase the degree of polarization is recovered. In particular, one can introduce a measure of the degree of polarization in a geometrical way as
\begin{equation}
    P(\rho) = \inf_{\sigma \in \U} d(\rho,\sigma),
    \label{eq:dopdistance}
\end{equation}
where $d(\rho,\sigma)$ is a distance or divergence that is contractive under the action of nonpolarizing operations, that is, $d(\Lambda(\rho),\Lambda(\sigma)) \leq d(\rho,\sigma)$.

Finally, let us note that any quantifier will establish a total order among the polarization states. However, as this total order is not intrinsic to the structure of $\P$, given any two polarized states, there may be different measures that assign contradictory values of the degree of polarization to them, that is, two measures can sort the states in a different way.

Let us apply this formalism for the cases of 2D and 3D random statistically stationary electromagnetic fields.
In each resource theory, we will use the same symbols $\U$ and $\prec$ to identify the set of unpolarized states and a hierarchy among the polarization states, respectively. Their meanings will be clear from the context.

\subsection{2D electromagnetic fields}
\label{subsec:2D}
For the 2D case, decomposition~\ref{eq:dp2D} is usually understood as a convex combination of two density matrices, $\rho^{\mathrm{1P}}$ and $\rho^{\mathrm{2U}}$, the first one representing the fully polarized state and the second one the completely unpolarized state (see, \eg,~\cite{WolfBook}). Accordingly, the unpolarized set is
\begin{equation}
    \U = \{\rho \in \P:  \rho = \rho^{\mathrm{2U}} \},
    \label{eq:unpolset2D}
\end{equation}
where $\rho^{\mathrm{2U}} = \frac{1}{2}\diag(1,1) = \frac{1}{2} I_2$. It is clear that the operations that preserve $\U$ are the unital ones, that is, transformations that satisfy $\Lambda\left(I_2/2\right) = I_2/2$~\cite{BengtssonBook}. These conditions can be posed in an equivalent way in terms of a majorization relation between $\rho$ and $\Lambda(\rho)$ (see, \eg,~\cite{MarshallBook}). More precisely, one has $\Lambda(\rho) \prec \rho$ iff $\Lambda$ is unital~\cite{Chefles2002}. In this case, the majorization relation $\Lambda(\rho) \prec \rho$ reduces to $\lambda_1 \leq \rho_1$, where $\rho_1$ and $\lambda_1$ are the greatest eigenvalues of $\rho$ and $\Lambda(\rho)$, respectively. Moreover, according to Uhlmann's theorem~\cite{Uhlmann1970}, one has $\Lambda(\rho) \prec \rho$ iff $\Lambda(\rho) = \sum_i p_i U_i \rho U_i^\dag$, where $p_i \geq 0$, $\sum_i p_i = 1$ and $\{U_i\}$ are $2 \times 2$ unitary matrices. In other words, nonpolarizing operations can be seen as random unitary transformations, always leading to a new state with a lower greatest eigenvalue.

It is important to note that for $d=2$ and for every pair $\rho, \sigma \in \P$, it is always true that either $\rho \prec \sigma$ or $\sigma \prec \rho$. Equivalently, for every pair $\rho, \sigma \in \P$, there exists a nonpolarizing operation $\Lambda$ consisting of a convex mixture of unitaries, as in Uhlmann's theorem, such that either $\rho = \Lambda(\sigma)$ or $\sigma = \Lambda(\rho)$. This leads to a natural total order among states, given by the greatest eigenvalue, which must be respected by any adequate 2D polarization monotone. As a consequence, any increasing function of the greatest eigenvalue is a polarization monotone.
Up to this point, there is no controversy and we recover the well-known results of the literature. For example, we have the degree of polarization for 2D fields~\cite{Wolf1959,AlQasimi2007},
\begin{equation}
    P^{2\mathrm{D}}(\rho) = \sqrt{1 - 4 \det \rho} = \rho_1 - \rho_2,
    \label{eq:dop2D}
\end{equation}
where $\det$ denotes the determinant. Indeed, one can see~\eqref{eq:dop2D} as the ratio of the average intensity of the polarized part of~\eqref{eq:dp2D} to the total averaged intensity of the field (assumed equal to $\Tr\rho=1$). Finally, we observe that~\eqref{eq:dop2D} can be rewritten in the form of~\eqref{eq:dopdistance} as
\begin{equation}
    P^{2\mathrm{D}}(\rho) = \|\rho- \rho^{\mathrm{2U}}\|_{\Tr},
    \label{eq:dop2Dgeo}
\end{equation}
where $\|\rho- \sigma \|_{\Tr} = \frac12\Tr|\rho-\sigma|$ is the trace distance between $\rho$ and $\sigma$.

\subsection{3D electromagnetic fields}
\label{subsec:3D}
Now, we are prepared to discuss the 3D case. As stated in Sec.~\ref{sec:introduction}, there is no consensus on how to define the unpolarized states for the 3D case. On the one hand, it is claimed that the unpolarized state should be invariant under arbitrary rotations of the field components and under arbitrary phase changes (see, \eg,~\cite{Setala2002a,Luis2005a}). On the other hand, it is claimed that an unpolarized state corresponds to a field which has no polarized component~\cite{Ellis2005a,Ellis2005b,Ellis2005c}. In the sequel, we formalize both situations from the resource-theoretic perspective, identifying the proper polarization monotones for each case.

\subsubsection{Resource theory based on majorization partial order}
\label{subsub:majorization}
Following~\cite{Luis2005a}, an unpolarized state has to be invariant under arbitrary rotations of the field components and under arbitrary phase changes. The only state that satisfies this is the one with equal eigenvalues, that is, $\rho_1 = \rho_2 = \rho_3 = \frac{1}{3}$~\cite{Setala2002a,Setala2002b}. In this way, the set of unpolarized states reduces to
\begin{equation}
    \U = \{\rho \in \P:  \rho = \rho^{\mathrm{3U}} \},
    \label{eq:unpolsetI}
\end{equation}
where $\rho^{\mathrm{3U}} = \frac{1}{3}\diag(1,1,1) = \frac{1}{3} I_3$, which can be seen as the natural generalization of the set given by~\eqref{eq:unpolset2D}. Again, the nonpolarizing operations are the unital ones that in this case satisfy
\begin{equation}
    \Lambda\left(\frac{I_3}{3}\right) = \frac{I_3}{3}.
    \label{eq:unital}
\end{equation}
This condition can be posed in an equivalent way in terms of a majorization relation,
\begin{equation}\label{eq:majorization}
  \Lambda(\rho) \prec \rho \quad \mbox{iff} \quad \lambda_1 \leq \rho_1 \ \mbox{and} \ \lambda_1+\lambda_2 \leq \rho_1+\rho_2,
\end{equation}
where $\{\rho_i\}$ and $\{\lambda_i\}$ are the eigenvalues of $\rho$ and $\Lambda(\rho)$, respectively, sorted in nonincreasing order. We have the same interpretation of $\Lambda$, as in the 2D case, as random unitary transformations. More precisely, one has
\begin{equation}\label{eq:randomunitary}
  \Lambda(\rho) \prec \rho \quad \mbox{iff} \quad \Lambda(\rho) = \sum_i p_i\, U_i\, \rho\, U_i^\dag,
\end{equation}
where $p_i \geq 0$, $\sum_i p_i = 1$ and $\{U_i\}$ are $3 \times 3$ unitary matrices. These random unitary transformations have already been analyzed in the study of irreversible behavior of the degree of polarization (see, \eg,~\cite{Refregier2008,Refregier2012}).

Unlike the 2D case, here there is not necessarily a majorization relation between any given pair of states $\rho, \sigma \in \P$. For instance, taking $\rho = \diag(0.6,0.2,0.2)$ and $\sigma = \diag(0.5,0.4,0.1)$, it is straightforward to check that neither $\rho \prec \sigma$ nor $\sigma \prec \rho$ is satisfied. Thus, we do not have a total order between the states. Majorization only provides a partial order. This means that for every $\rho, \sigma, \omega \in \P$ one has (i)  $\rho \prec \rho$ (reflexivity), (ii) if $\rho \prec \sigma$ and $\sigma \prec \rho$, then $\rho = \sigma$ (antisymmetry), and (iii) if $\rho \prec \sigma$ and $\sigma \prec \omega$, then $\rho \prec \omega$ (transitivity). A criterion to compare polarization states for the three-dimensional case by means of this partial order has been recently introduced in~\cite{Gamel2014}.

The polarization monotones within this resource theory are given by Schur-convex functions, that is, functions that preserve the majorization relation: if $\rho \prec \sigma$, then $P(\rho) \leq P(\sigma)$. Among all Schur-convex functions, we will see that some well-known measures of degree of polarization introduced in the literature can be rewritten as in the form given by~\eqref{eq:dopdistance}.

Let us first consider the degree of polarization of Set\"al\"a-Shevchenko-Kaivola-Friberg based on the purity of $\rho$~\cite{Setala2002a},
\begin{equation}
    P^{\SSKF}(\rho) = \sqrt{\frac{3}{2}\left(\Tr \rho^2- \frac{1}{3}\right)}.
    \label{eq:SSKF}
\end{equation}
It is straightforward to check that this measure satisfies the Schur-concave condition. This quantity can be rewritten in several equivalent ways, for instance, in terms of a generalization of Stokes parameters for three-dimensional fields (see, \eg,~\cite{Barakat1977,Barakat1983,Carozzi2000,Setala2002a,Setala2002b,Luis2005b}). In particular, $P^{\SSKF}(\rho)$ has a clear geometric interpretation as the Hilbert-Schmidt distance between $\rho$ and the unpolarized state $\rho^{\mathrm{3U}}$ (see, \eg,~\cite{Luis2005a}), that is,
\begin{equation}
    P^{\SSKF}(\rho) = \sqrt{\frac{3}{2}} \|\rho - \rho^{\mathrm{3U}}\|_{\HS},
    \label{eq:HS-SSKF}
\end{equation}
where $\|\rho - \sigma\|_{\HS} = \sqrt{\Tr (\rho - \sigma)^2}$ is the Hilbert-Schmidt distance between $\rho$ and $\sigma$.

Another interesting polarization monotone is based on the von Neumann entropy of the state~\cite{Dennis2007,Gamel2012},
\begin{equation}
    P^{\vN}(\rho) = 1 - \frac{S(\rho)}{\ln 3},
    \label{eq:VN}
\end{equation}
where $S(\rho) = - \Tr \rho \ln \rho$ is the von Neumann entropy. The measure~\eqref{eq:VN} also has a geometric interpretation as the normalized relative entropy between $\rho$ and $\rho^{\mathrm{3U}}$, that is,
\begin{equation}
    P^{\vN}(\rho) = \frac{1}{\ln 3} S(\rho \| \rho^{\mathrm{3U}}),
    \label{eq:VNgeo}
\end{equation}
where $S(\rho\| \sigma) = \Tr[ \rho (\ln \rho - \ln \sigma)]$ is the relative entropy (or quantum divergence) between $\rho$ and $\sigma$. Let us note that the von Neumann entropy is a particular case of the Schur-concave generalized $(h,\phi)$ entropies~\cite{Bosyk2016}, so that it is feasible to extend the measure in~\eqref{eq:VN} by appealing to this family of generalized entropies as follows:
\begin{equation}\label{eq:hphient}
  P^{(h,\phi)}(\rho) = 1 - \frac{S_{(h,\phi)}(\rho)}{h\left(3 \phi\left( \frac{1}{3} \right)\right)},
\end{equation}
where $S_{(h,\phi)} = h(\Tr \phi(\rho))$ and the entropic functionals are such that $h : \mathbb{R} \rightarrow \mathbb{R}$ and $\phi: [0, 1]  \rightarrow \mathbb{R}$, with an increasing $h$ and a concave $\phi$, or  a decreasing $h$ and a convex $\phi$; and the additional conditions  $\phi(0) = 0$ and $h(\phi(1)) = 0$. Let us also note that the degree of polarization~\eqref{eq:SSKF} can be expressed in an entropic form as $P^{\SSKF}(\rho) = \sqrt{P^{(1-x,x^2)}(\rho)} = \sqrt{1 - \frac{3}{2}S_{(1-x,x^2)}(\rho)}$.

Furthermore, we can consider a quantifier that is linear with respect to the difference of the greatest and lowest eigenvalues~\cite{Dennis2007},
\begin{equation}
    P^{\lin}(\rho) = \rho_1 - \rho_3.
    \label{eq:lin}
\end{equation}
It has been proven that $P^{\lin}(\rho)$ is Schur-convex~\cite{Gamel2014}. In addition, this monotone can be expressed in the form~\eqref{eq:dopdistance} as
\begin{equation}
    P^{\lin}(\rho) = d(\rho,\rho^{\mathrm{3U}}),
    \label{eq:lin2}
\end{equation}
where $d(\rho,\sigma)= |\rho_1 -\sigma_1|+|\rho_3 -\sigma_3|$ is a proper distance between $\rho, \sigma \in \P$.

Finally, let us note that the results of this resource theory based on majorization partial order are in concordance with the ones given in~\cite{Refregier2012, Gamel2014}. In~\cite{Refregier2012}, the author studies the nonincreasing property of 3D degrees of polarization under random unitary transformations. However, we remark that a discussion about these transformations in connection with majorization theory, unital transformations, and Schur-convex functions was not provided. On the other hand, as we have already noticed, a majorization criterion applied to the polarization density matrices has also been introduced in~\cite{Gamel2014}, but its motivation differs from our approach.

\subsubsection{Resource theory based on convex preorder}
\label{subsub:convex}
We now adopt the viewpoint proposed in~\cite{Ellis2005a,Ellis2005b,Ellis2005c}, where it is asserted that an unpolarized state corresponds to a field which has no polarized component. This happens when the two greatest eigenvalues of $\rho$ are equal: $\rho_1 = \rho_2$. From decomposition~\eqref{eq:dp3D}, we see that the corresponding set of unpolarized states is the convex set
\begin{equation}
    \U = \{\rho \in \P: \rho = p \rho^{\mathrm{2U}} + (1-p) \rho^{\mathrm{3U}} \;\mbox{with}\, p \in [0,1]\}.
    \label{eq:unpolsetII}
\end{equation}
Unlike the previous case, now $\U$ is not just a single state but an entire convex subspace determined by the segment joining $\rho^{\mathrm{2U}}$ with $\rho^{\mathrm{3U}}$ [see Fig.~\ref{fig:polspace}(b)]. In order to provide a suitable resource theory, we must identify the class of operations that preserves this $\U$. A rather natural option for those nonpolarizing operations, $\Lambda$, is given by the operations that involve mixing with a member of $\U$, that is,
\begin{equation}
    \Lambda(\rho) = p\,\rho + (1-p)\,\omega, \;\text{with}\; p \in [0,1] \;\text{and}\; \omega\in\U.
    \label{eq:unpolopII}
\end{equation}
Due to the convexity of $\U$, this class of operations given by~\eqref{eq:unpolopII} for any $\omega\in\U$ preserves the unpolarized set. Based on the work by Sperling and Vogel~\cite{Sperling2015}, where the authors show how to define a convex preorder for quantum states with respect to an arbitrary convex set of states, we can see that the class of operations of~\eqref{eq:unpolopII} induces a preorder relation (a binary relation that is  reflexive and transitive, but is not necessarily symmetric as in the case of a partial order), denoted by $\prec$, for any two polarization states in the following manner:
\begin{equation}\label{eq:convexpreorder}
  \rho \prec \sigma \quad \mbox{iff} \quad \exists \, \Lambda \ \mbox{such that} \ \rho =  \Lambda(\sigma),
\end{equation}
where $\Lambda$ accounts for any nonpolarizing operation of the form~\eqref{eq:unpolopII}. This convex preorder captures the idea that a given state $\rho$ has a lower degree of polarization than $\sigma$ whenever the former is obtainable as a convex combination between the latter and any unpolarized state. This preorder is indeed a partial order for the set $\P \backslash \U$. Although $\rho \prec \sigma$ and $\sigma \prec \rho$ are satisfied for any pair of $\rho,\sigma \in \U$, this does not necessarily imply $\rho = \sigma$. However, we can consider all states in $\U$ as equivalent.

We now intend to look for the adequate measures that behave monotonically with respect to the class of operations defined in~\eqref{eq:unpolopII}. We do know that the polarization monotones of our previous case, namely, $P^{\SSKF}, P^{\vN}$, and $P^{\lin}$, do not work for this new prescription of the unpolarized set since those monotones do not put $\rho^{\mathrm{2U}}$ and $\rho^{\mathrm{3U}}$ on equal footing. We prove now that a measure of the degree of polarization of Ellis-Dogariu-Ponomarenko-Wolf, defined as the difference between the greatest eigenvalues, that is~\cite{Ellis2005b},
\begin{equation}
    P^{\mathrm{EDPW}}(\rho) = \rho_1 - \rho_2,
    \label{eq:polEDPW}
\end{equation}
fits well in this second approach. First, given that the unpolarized states of~\eqref{eq:unpolsetII} are those with $\rho_1=\rho_2$, we have that $P^{\mathrm{EDPW}}(\rho)=0$ iff $\rho\in\U$. In addition, one has that this polarization monotone is bounded: $0 \leq P^{\mathrm{EDPW}}(\rho) \leq 1$. Finally, it is direct to see that $P^{\mathrm{EDPW}}(\Lambda(\rho))  = p \, (\rho_1 - \rho_2 ) = p \, P^{\mathrm{EDPW}}(\rho) \leq P^{\mathrm{EDPW}}(\rho)$ with $p \in [0,1]$, which is the monotonicity condition. Interestingly enough, we observe that $P^{\mathrm{EDPW}}$ can also be expressed as a distance to the set of unpolarized states $\U$ by means of the trace distance, that is,
\begin{equation}
    P^{\mathrm{EDPW}} = \min_{\sigma\in\U} \|\rho-\sigma\|_{\Tr}.
    \label{eq:EDPWgeo}
\end{equation}

Our approach allows one to see that another suitable polarization monotone, in this case, is clearly given by $P^{\mathrm{RE}}(\rho) = \min_{\sigma\in\U} S(\rho \| \sigma)$.

\subsubsection{Comparisons between both resource theories}
\label{subsub:comparison}
First, in both resource theories the state $\rho^\mathrm{1P}$ is the unique fully polarized state, and any other state $\rho$ can be obtained from $\rho^\mathrm{1P}$ by means of the corresponding nonpolarizing operations: $\rho= \Lambda\left(\rho^\mathrm{1P}\right)$, where $\Lambda$ is given by~\eqref{eq:unital} or~\eqref{eq:unpolopII}, respectively. In other words, $\rho \prec \rho^\mathrm{1P}$ for all $\rho \in \P$ where $\prec$ is majorization or the convex preorder, respectively. As a consequence, all the polarization monotones assign the maximum value to $\rho^\mathrm{1P}$, that is, $P^X(\rho) = 1$ iff $\rho =\rho^\mathrm{1P}$ for $X=\SSKF,\vN,\lin,(h,\phi),$ or $\EDPW$.

In spite of the previous fact, both proposals have more differences than similarities. Indeed, they are built onto different notions for the unpolarized set and the nonpolarizing operations. Therefore, different hierarchies for the polarized states arise, respectively, given by the majorization partial order and the convex preorder (see Table~\ref{tab:monotones3D}).

\begin{table}[htbp]
\vspace{-4mm}
\centering
\caption{\bf Resource theories for 3D random fields}
\begin{tabular}{ccc}
\hline
 unpolarized states ($\U$) & order ($\prec$) & monotones ($P$) \\
\hline
  $\rho^\mathrm{3U}$ &majorization & $P^\SSKF, P^\vN, P^\lin$ \\
 $p\rho^\mathrm{2U}+(1-p)\rho^\mathrm{3U}$ & convex mixing  & $P^\EDPW$ \\
\hline
\end{tabular}
  \label{tab:monotones3D}
\end{table}

\begin{figure}[htbp]
 \centering
 \includegraphics[width=\linewidth]{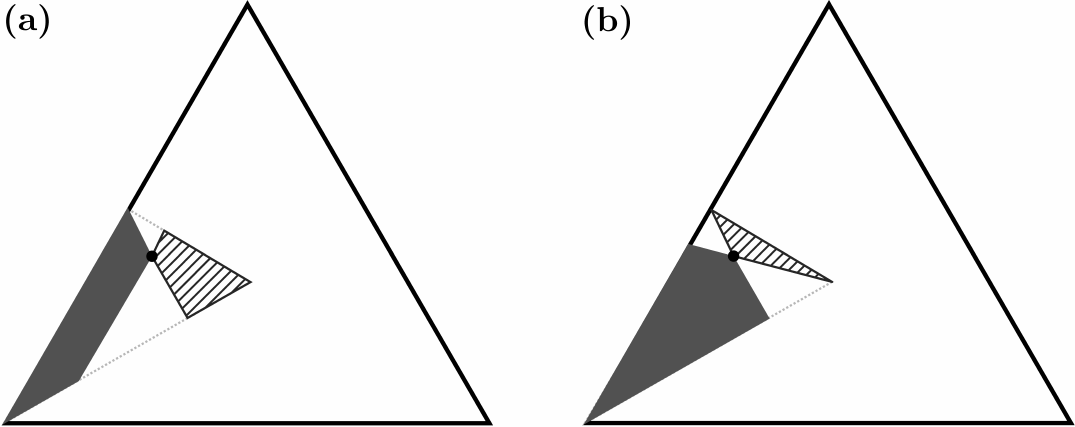}
 \caption{Given the state $\rho = \diag(0.5,0.4,0.1)$ (black point), we depict the sets $\rho^{\prec}$ (gray region), $\rho^{\succ}$ (meshed region), and $\rho^\mathrm{inc}$ (white region) given by (a) majorization partial order, and (b) convex preorder.}
 \label{fig:orders}
\end{figure}

In order to visualize this fact, let us introduce the following sets:
for a given $\rho$, let $\rho^\prec$ be the set of all states with greater degree of polarization, $\rho^\succ$ be the set of all states with lower degree of polarization, and let $\rho^\mathrm{inc}$ be the set of all the incomparable states. More precisely, $\rho^{\prec} =\{\sigma \in \P: \ \rho \prec \sigma \}$, $\rho^{\succ} =\{\sigma \in \P: \  \sigma \prec \rho \}$, and $\rho^{\mathrm{inc}} =\{\sigma \in \P: \ \rho \not\prec \sigma \ \text{and} \ \sigma \not\prec \rho\}$, where $\prec$ indicates majorization relation or convex preorder depending on the resource theory considered. In Fig.~\ref{fig:orders}, we illustrate these sets for a given polarization density matrix, $\rho = \diag(0.5,0.4,0.1)$. The gray, meshed, and white regions represent the sets $\rho^{\prec}$, $\rho^{\succ}$, and $\rho^{\mathrm{inc}}$, respectively, which are different in each theory. However, as expected, in both cases the state $\rho^{1\mathrm{P}}$ belongs to the gray region and the corresponding set of unpolarized states is included in the meshed one.

\begin{figure}[htbp]
 \centering
 \includegraphics[width=\linewidth]{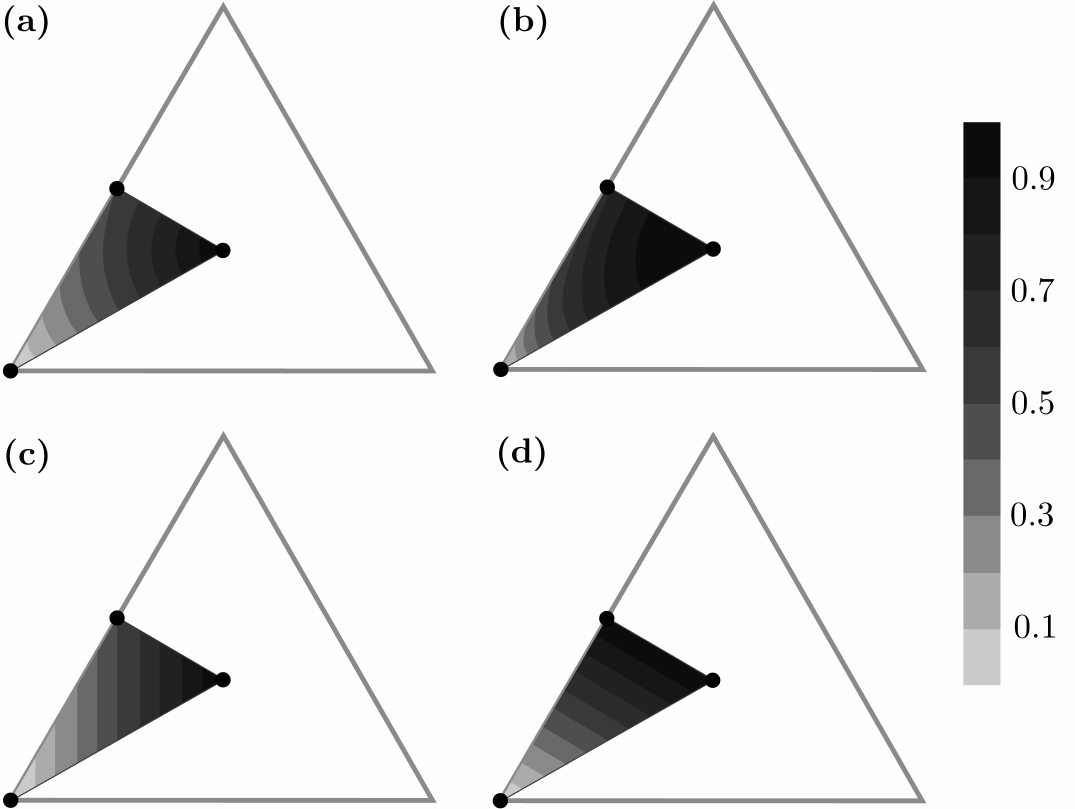}
 \caption{Geometric representation of the isopolarization curves for (a) $P^\SSKF$, (b) $P^\vN$, (c) $P^\lin$, and (d) $P^\EDPW$. All of them monotonically increase from the center towards the vertex.}
 \label{fig:isopol}
\end{figure}

Finally, in Fig.~\ref{fig:isopol}, we show how the 3D polarization monotones behave, depicting the contour plots of isopolarization curves for each polarization monotone, namely, $P^X(\rho) = c$ with $c \in [0,1]$. For the resource theory based on majorization, the measures increase as they move away from the state $\rho^{3\mathrm{U}}$. In particular, $P^\SSKF$ and $P^\lin$ behave similarly; indeed they assign the same value of $0.5$ to the degree of polarization of the state $\rho^{2\mathrm{U}}$. Regarding the monotone $P^\EDPW$, we see that the contours are parallel to the segment that represent the unpolarized set and they increase as they move away from this segment.

\section{Concluding remarks}
\label{sec:conclusions}
In summary, we have demonstrated the powerfulness of a resource-theory formulation of polarization that merges, in a single framework, different and even conflicting polarization measures. This endows polarization with rather fruitful and sounded order-theoretic structures, such as majorization and convex mixing, derived from transformation properties. In addition, this allows us to reformulate in a geometrical way the best-known existing 2D and 3D degrees of polarization, as a minimum distance (or divergence) from the measured state to a set of unpolarized states, which is closed under the corresponding class of nonpolarizing operations. These results are timely and relevant since they put polarization at the level of other resources for modern information technologies, such as quantum coherence and entanglement. This is consistent as far as polarization is a form of coherence, characterized by robustness and an extremely simple experimental implementation.

\section*{Acknowledgement}
G.M.B. and G.B. acknowledge CONICET (Argentina) for financial support. AL acknowledges financial support from the Spanish Ministerio de Economía y Competitividad, Project No. FIS2016-75199-P, and from the Comunidad Autónoma de Madrid research consortium QUITEMAD+, Grant No. S2013/ICE-2801.

\bibliography{arXivBosykBellomoLuis}

\end{document}